\documentclass{article}

\PassOptionsToPackage{numbers, compress}{natbib}

\usepackage[final]{neurips_2019}

\usepackage[utf8]{inputenc} 
\usepackage[T1]{fontenc}    
\usepackage{hyperref}       
\usepackage{url}            
\usepackage{booktabs}       
\usepackage{amsfonts}       
\usepackage{nicefrac}       
\usepackage{microtype}      
\usepackage{amssymb}
\usepackage{bm}
\usepackage{dsfont}
\usepackage[noend]{algpseudocode}
\usepackage{dsfont}
\usepackage{subeqnarray}
\usepackage{amsmath,graphicx}
\usepackage[normalem]{ulem}
\usepackage{amsthm}

\newtheorem{assumption}{Assumption}
\newenvironment{proof-sketch}{\noindent{\bf Sketch of Proof}\hspace*{1em}}{\qed\bigskip}

\usepackage{float}
\usepackage{appendix}
\usepackage{comment}
\usepackage{enumitem}
\usepackage{algorithm}
\usepackage[noend]{algpseudocode}
\newcommand{\RR}{{\mathbb{R}}}
\usepackage{multirow}
\usepackage[format=plain,justification=raggedright,singlelinecheck=false]{caption}
\usepackage{xr}
\newtheorem{remark}{Remark}
\newtheorem{claim}{Claim}

\newtheorem{method}{Method}	

\title{Revisiting the Bethe-Hessian: Improved Community Detection in Sparse Heterogeneous Graphs}

\author{%
	Lorenzo Dall'Amico \\
	GIPSA-lab, UGA, CNRS, Grenoble INP\\
	lorenzo.dall-amico@gipsa-lab.fr
	\AND
	Romain Couillet \\
	GIPSA-lab, UGA, CNRS, Grenoble INP\\
	L2S, CentraleSup\'elec, University of Paris Saclay\\
	\And
	Nicolas Tremblay \\
	GIPSA-lab, UGA, CNRS, Grenoble INP\\
}

\begin{document}
	
	\maketitle
	
	\begin{abstract}
		Spectral clustering is one of the most popular, yet still incompletely understood, methods for community detection on graphs. This article studies spectral clustering based on the Bethe-Hessian matrix $H_r = (r^2-1)I_n + D-rA$ for sparse heterogeneous graphs (following the degree-corrected stochastic block model) in a two-class setting. For a specific value $r = \zeta$, clustering is shown to be insensitive to the degree heterogeneity. We then study the behavior of the informative eigenvector of $H_{\zeta}$ and, as a result, predict the clustering accuracy. The article concludes with an overview of the generalization to more than two classes along with extensive simulations on synthetic and real networks corroborating our findings.
	\end{abstract}
	
	\section{Introduction}
	
	Network theory studies the interaction of connected systems of agents. Real networks tend to be structured in affinity classes and the problem of clustering consists in retrieving these unknown classes from the observed network pairwise interactions \cite{fortunato2010community}.
	Belief propagation (BP) is an efficient way to reconstruct communities and  -- under certain conditions (see \cite{mossel2014belief}) -- was proved to give \emph{optimal} reconstruction. On the negative side, BP suffers from a possibly long convergence time and a  non-trivial implementation. Among the alternative clustering algorithms, spectral techniques proved particularly efficient in terms of speed and analytical tractability \cite{rohe2011spectral,von2007tutorial,gulikers2017spectral,lei_consistency_2015}. In the dense regime, in particular, 
	where the average node degree scales like the size of the network, random matrix theory \cite{von2007tutorial,nadakuditi2012graph,ali2016random} manages to predict the asymptotic spectral clustering performances and to identify transition points beyond which asymptotic non trivial classification is achievable. This is however not the typical condition for real networks that tend instead to be \emph{sparse}. For a graph $\mathcal{G}(\mathcal{V},\mathcal{E})$ with $|\mathcal{V}|=n$ nodes, the condition of sparsity means that the average degree $d$ does not depend on the size of the network and in particular $d \ll n$. 
	
	Both standard spectral clustering methods and their associated random matrix asymptotics collapse in this regime. As an answer, many intuitions emerged from statistical physics and led to important seminal steps. Notably, two deeply connected matrices recently proved to overcome the problem of sparsity: the $n \times n$ Bethe-Hessian \cite{saade2014spectral} $H_r$ with $r \in \mathbb{R}$ a parameter to be fixed -- the study of which is the object of the present article--, and the non symmetric non backtracking operator $B \in \{0,1\}^{2|\mathcal E|\times2|\mathcal E|}$ \cite{krzakala2013spectral}. Both matrices were introduced and studied under the homogeneous degree stochastic block model (SBM). Narrowing to the case of two communities it was proved both experimentally and theoretically \cite{bordenave2015non,mossel2014belief,massoulie2014community,mossel2015reconstruction} that, if there exists an algorithm able to detect communities better then random guess, then these two matrices can be used to give non-trivial node partition. It is said that both algorithms work \emph{down to the detectability threshold}.
	
	However, real networks are rarely homogeneous and typically follow a power law degree distribution \cite{barabasi1999emergence}. The results of \cite{gulikers2016non,gulikers2015impossibility} generalize the above studies to heterogeneous networks, generated by degree-corrected stochastic block models (DC-SBM) \cite{karrer2011stochastic} and suggest that both $B$ and $H_r$ provide also in this case non trivial clustering down to the detectability threshold. Yet, a precise characterization of their behavior and performances is still lacking; the present article shows that some aspects of the behavior of $B$ and $H_r$ have indeed been overlooked.

	Spectral clustering in sparse heterogeneous networks has also been tackled using various regularized Laplacian matrices \cite{qin2013regularized,le2017concentration,joseph2013impact} but, to our knowledge, these are not proved to operate down to the detectability threshold. These structurally different methods are discussed in concluding remarks.
	
	The main message of the present communication is that, under a DC-SBM setting, the choice of $r$ in $H_r$ proposed in \cite{saade2014spectral} for the SBM setting is suboptimal. We propose and theoretically support an improved parametrization $r=\zeta$ that allows the Bethe-Hessian $H_\zeta$ to efficiently detect communities in sparse and heterogeneous graphs. In detail, under the DC-SBM setting, a) we propose a spectral algorithm on $H_\zeta$ which performs efficiently down to the detectability threshold, with an informative eigenvector not tainted by the degree distribution (unlike in \cite{saade2014spectral}); b) 
	the algorithm is generalized to $k$-class clustering with a consistent estimation procedure for $k$; c) substantial performance improvements on the originally proposed Bethe-Hessian are testified by simulations on synthetic and real networks.
	
	The remainder of the article is organized as follows: Section~\ref{sec:choice} argues on the optimal value $r = \zeta$ for $H_r$ and, based on heuristic arguments, studies the behavior of the informative eigenvector of $H_\zeta$, concluding with an explicit expression of the clustering performance; Section~\ref{sec:estimate} provides an unsupervised method to estimate $\zeta$, drawing on connections with the non-backtracking matrix $B$; Section~\ref{sec:extension} extends the algorithm to a $k$-class scenario; numerical supports are then provided in Section~\ref{sec:validation} on both synthetic and real networks; concluding remarks close the article.
	
	\smallskip
	
	\noindent{\bf Reproducibility.} 
	A Python implementation of the proposed algorithm along with codes to reproduce the results of the article are available at \href{http://lorenzodallamico.github.io/codes}{lorenzodallamico.github.io/codes}.
	
	\section{Model and Main Results}
	\label{sec:choice}
	
	\subsection{Model setting}
	
	Consider an undirected binary graph $\mathcal{G}(\mathcal{E},\mathcal{V})$, with nodes $\mathcal{V}=\{1,\ldots,n\}$ ($|\mathcal{V}|=n$) and edges $\mathcal{E}\subset \mathcal{V}\times \mathcal{V}$ ($|\mathcal{E}|=m$). Let $\bm{\sigma} \in \{-1,1\}^n$ be the vector of class labels, both classes being of equal size (i.e., $\sum_i\sigma_i=0$), and $C=(\begin{smallmatrix} c_{\rm in} & c_{\rm out} \\ c_{\rm out} & c_{\rm in}\end{smallmatrix})$. These assumptions are meant to set the problem in a more readable symmetric scenario. Section~\ref{sec:extension} extends the results to multiple classes of possibly different sizes.
	In order to account both for sparsity and heterogeneity, we consider the DC-SBM as a generative model for $\mathcal{G}$. Denoting $A \in \{0,1\}^{n \times n}$ the adjacency matrix defined by $A_{ij} = 1_{(i,j)\in\mathcal{E}}$, the DC-SBM generates edges independently according to:
	\begin{equation}
	\mathbb{P}(A_{ij}=1|\sigma_i,\sigma_j,\theta_i,\theta_j) = \theta_i\theta_j\frac{C_{\sigma_i,\sigma_j}}{n},
	\label{eq:DCSBM}
	\end{equation}
	where $\bm{\theta}=(\theta_1,\ldots,\theta_n)$ is the vector of random intrinsic connection ``probabilities'' of each node.
	The $\theta_i$'s are assumed i.i.d.\@ and independent of the class labels, and we impose $\mathbb{E}[\theta_i]=1$, $\mathbb{E}[\theta_i^2]=\Phi$. The $1/n$ term bounds the degree of each node to an $n$-independent value, making the network sparse.
	Denoting  $c = (c_{\rm in } + c_{\rm out})/2$, the detectability condition \cite{gulikers2015impossibility} reads:
	\begin{equation}
	\alpha \equiv \frac{c_{\rm in} - c_{\rm out}}{\sqrt{c}} \geq \frac{2}{\sqrt{\Phi}} \equiv \alpha_{\rm c}.
	\label{eq:dec_th}
	\end{equation}
	For $\alpha < \alpha_c$, no algorithm can partition the nodes better than by random guess. 
	Letting $D = {\rm diag}(A\mathds{1})$ be the degree matrix, the Bethe-Hessian is defined as
	\begin{equation}
	H_r = (r^2-1)I_n + D - rA, \quad r \in \RR.
	\end{equation}
	This matrix was originally proposed in \cite{saade2014spectral} for $r = \sqrt{c\Phi}$, which asymptotically provides non trivial clustering down to the \emph{detectability threshold} (for $\alpha > \alpha_c$). The informative eigenvector of $H_r$ is associated with the second smallest eigenvalue and we denote it $\bm{x}^{(2)}_r$. The components of $\bm{x}_{\sqrt{c\Phi}}^{(2)}$ are however strongly tainted by the $\theta_i$'s, sensibly altering the algorithm performance.
	
	We show here that for $\alpha \geq \alpha_c$ there exists a value  $\zeta \leq \sqrt{c\Phi}$ for which the components of the second eigenvector $\bm{x}_{\zeta}^{(2)}$ of $H_\zeta$ align to the labels irrespective of the $\theta_i$'s, thus largely improving the algorithm performance while maintaining detectability down to the threshold.

	\subsection{Informative eigenvector of $H_r$}
	
	In the sequel we assume that: (i) being sparse, we can locally approximate the graph by a tree
	\cite{dembo2010gibbs} and therefore $\mathbb{P}(\bm{\sigma}_{\partial_i}|\sigma_i) \simeq \prod_{j \in \partial_i} \mathbb{P}(\sigma_j|\sigma_i)$, with $\partial_i$ the neighbourhood of $i$;
	(ii) $n \to \infty$ and $c$ is bounded by an $n$-independent value while being arbitrarily larger than one, i.e., $n \gg c \gg 1$.
	
	
	For ease of notation we work here with $D - rA$ rather than $H_r$, both having the same eigenvectors. The core of our proposed method lies in the following observation, related to the action of $H_r$ on $\bm\sigma$:
	\begin{equation}
	[(D - rA)\bm{\sigma}]_i =  d_i\sigma_i\left[1 - r\left(\frac{|\partial_i^{(s)}|}{d_i}-\frac{|\partial_i^{(o)}|}{d_i}\right)\right]
	\label{eq:D-rAsigma}
	\end{equation}
	where $|\partial_i^{(s)}|$ (resp., $|\partial_i^{(o)}|$) stands for the number of neighbors of $i$ belonging to the same (resp., opposite) class as $i$. 
	We show next that a proper choice of $r$ can annihilate the right-hand side of \eqref{eq:D-rAsigma} ``on average'' or whenever the typical degrees $d_i$ are not too small, turning \eqref{eq:D-rAsigma} into an eigenvector equation. To this end, we need to quantify the random variables $|\partial_i^{(s)}|$ and $|\partial_i^{(o)}|$.
	
	From a Bayesian perspective, $\bm{\sigma}$ and $\bm{\theta}$ are unknown parameters and $A$ (and thus $d_i$) known realizations. We may thus write 
	\begin{align*}
	\mathbb{P}(\sigma_i|\sigma_j,A_{ij} = 1) &= \frac{\mathbb{P}(\sigma_i,\sigma_j|A_{ij}=1)}{\mathbb{P}(\sigma_j|A_{ij}=1)} = 2\iint \mathbb{P}(\sigma_i,\sigma_j,\theta_i,\theta_j|A_{ij} = 1) d\theta_i d\theta_j\\
	&\propto \iint \mathbb{P}(A_{ij} = 1|\sigma_i,\sigma_j,\theta_i,\theta_j)\mathbb{P}(\sigma_i,\sigma_j,\theta_i,\theta_j)d\theta_i d\theta_j \propto C(\sigma_i,\sigma_j),
	\end{align*}
	where we used the facts that the classes are of equal size ($\mathbb{P}(\sigma_i)$  is constant), and the $\theta_i$ are i.i.d., independent of the classes with $\mathbb{E}[\theta_i]=1$. Normalizing, one finally obtains $\mathbb{P}(\sigma_i|\sigma_j,A_{ij} = 1)=C(\sigma_i,\sigma_j)/(c_{\rm in} + c_{\rm out})$, which is independent of the degree distribution. We further know that $|\partial_i^{(s)}|+|\partial_i^{(o)}| = d_i$, which is a deterministic observation. Given the locally tree-like structure of the graph, neighbors of the same node are conditionally independent -- see (i) -- so that $|\partial_i^{(s)}|$ is the sum of $d_i$ i.i.d.\@ Bernoulli random variables with parameter $p = c_{\rm in}/(c_{\rm in}+c_{\rm out})$. We thus obtain
	\begin{equation}
	\label{eq:D-rA_expectation}
	\mathbb{E}[[(D-rA)\bm{\sigma}]_i~|~A] = d_i\sigma_i \left(1-r\frac{c_{\rm in}-c_{\rm out}}{c_{\rm in}+c_{\rm out}}\right).
	\end{equation}
	This equation suggests that, for the expectation of \eqref{eq:D-rAsigma} to be an eigenvector equation in the large (but finite) $d_i$ regime, $r$ should be taken equal to
	\begin{align}
	\label{eq:def_zeta}
	r = \frac{c_{\rm in}+c_{\rm out}}{c_{\rm in}-c_{\rm out}} = \frac{2\sqrt{c}}{\alpha} \equiv \zeta_{\alpha} .
	\end{align}
	with $\alpha$ as in \eqref{eq:dec_th} the proper control parameter 
	for the clustering problem (as shown e.g., in \cite{nadakuditi2012graph,gulikers2016non,gulikers2015impossibility,decelle2011asymptotic}).
	For simplicity of notation the dependence on $\alpha$ of $\zeta=\zeta_\alpha$ will be made explicit only when relevant. 
	Intuitively, this calculus suggests that $\zeta$ is the only value of $r$ that ensures that $H_r$ has an informative eigenvector not significantly tainted by the degree distribution. This claim is supported by the following two remarks. 
	
	\begin{remark}[Consistency of $\zeta $ for trivial classification]
		\label{rem:consistency}
		In the limit of trivial clustering where $c_{\rm out} \to 0$, $|\partial_i^{(s)}|$ and $|\partial_i^{(o)}|$ tend to their mean. In particular, for $c_{\rm out}=0$, $\zeta =1$ and $(D-\zeta A)\bm{\sigma} = (D-A)\bm{\sigma}= 0$, so that $\bm{\sigma}$ is an exact eigenvector of $H_{\zeta=1}$ associated with its zero eigenvalue.
	\end{remark}
	
	\begin{remark}[Mapping to Ising]
		\label{rem:B_opt}
		The original intuition behind the Bethe-Hessian matrix arises from a mapping of the community labels into the spins of a Ising Hamiltonian. The ``temperature-related'' parameter $r$ guarantees a correct mapping \emph{only} for $r = \zeta$. This is elaborated in details in Section~\ref{app:map to ising} of the supplementary material.
	\end{remark}
	
	Although one commonly assumes an assortative model for the communities, by which $c_{\rm in}>c_{\rm out}$, the Bethe-Hessian matrix is oblivious of the sign of $c_{\rm in}-c_{\rm out}$. 
	\begin{remark}[Disassortative networks]
		\label{rem:dis}
		The case where $c_{\rm out} > c_{\rm in}$ does not invalidate the above analysis which results in $\zeta < 0$. Clustering with $H_\zeta$ is thus also valid in disassortative networks.
	\end{remark}
	
	In practice, for a given (non averaged) realization of the $\sigma_i$'s, 
	$\bm{\sigma}$ is not an exact eigenvector of $H_\zeta$. By a perturbation analysis around $\bm{\sigma}$, we next analyze the behavior of the corresponding informative eigenvector of $H_\zeta$ and theoretically predict the overlap performance. 
	
	\subsection{Performance Analysis}
	\label{sec:insights}
	
	To generalize the averaged analysis of \eqref{eq:D-rA_expectation}, we perturb $\bm\sigma$ by a ``noise'' term $\bm{\delta}$ and write $\bm{x}_{\zeta}^{(2)} \equiv \bm{\sigma} + \bm{\delta}$. Since $\zeta$ is however maintained, the associated eigenvalue of $D-\zeta A$, which is zero in \eqref{eq:D-rA_expectation}, now possibly deviates from zero; this eigenvalue is denoted $\lambda_\alpha$, i.e.,
	\begin{equation}
	\label{eq:eigenvalue_eq}
	(D-\zeta_{\alpha} A)(\bm{\sigma}+\bm{\delta}) = \lambda_{\alpha}(\bm{\sigma}+\bm{\delta}).
	\end{equation}
	From Remark~\ref{rem:consistency}, we already know that $\lim_{\alpha \to \sqrt{2c_{\rm in}}} \lambda_\alpha = 0$. 
	
	
	In the following, expectations are taken for a fixed realization of the network, \emph{i.e.} $\mathbb{E}[\cdot] \equiv \mathbb{E}[\cdot | A]$. Writing $|\partial_i^{s}| = \mathbb{E}[|\partial_i^{s}| ] + \Delta_i$ and $|\partial_i^{o}| = \mathbb{E}[|\partial_i^{o}|] - \Delta_i$, where we exploited the relation $|\partial_i^{s}|+|\partial_i^{o}|=d_i$, we obtain:
	\begin{equation}
	[(D-\zeta_{\alpha} A)(\bm{\sigma}+\bm{\delta})]_i = -2\zeta_{\alpha} \sigma_i \Delta_i + d_i\delta_i - \zeta_{\alpha}  \sum_{j \in \partial_i}\delta_j.
	\label{eq:noise}
	\end{equation}
	The random variable $\Delta_i$ is a sum of $d_i$ independent  (centered) Bernoulli random variables, tending in the large $c$ limit to a zero mean Gaussian, i.e.,
	\begin{align}
	\Delta_i \sim \mathcal{N}(0,d_ic_{\rm in}c_{\rm out}/(c_{\rm in}+c_{\rm out})^2) \equiv \mathcal{N}(0,d_i f^2_{\alpha}/\zeta_{\alpha} ^2),\quad f_{\alpha} \equiv \frac{\sqrt{c_{\rm in}c_{\rm out}}}{c_{\rm in}-c_{\rm out}} = \frac1\alpha \sqrt{c-\frac{\alpha^2}{4}}.
	\label{eq:f_alpha}
	\end{align}

	Our analysis of \eqref{eq:noise} relies on the following claim 
	that we shall justify next.
	\begin{assumption}
		\label{ass:delta_i}
		The random variables $\delta_i$, $1\leq i\leq n$, are distributed as $\delta_i \sim \mathcal{N}(-\mu_{\alpha} \sigma_i, f^2_{\alpha}\beta_i^2) $
		for some $\mu_\alpha\in\RR$ depending on $\alpha$ only, and $\beta_i\in\RR$ depending on $i$ only. Besides, the $\delta_i$'s are ``weakly dependent'' in the sense that $\mathbb{E}[\delta_i\delta_j]=\mathbb{E}[\delta_i]\mathbb{E}[\delta_j]+O(1/c)$.
	\end{assumption}
	
	The elements of Assumption~\ref{ass:delta_i} rely on the following observations:
	\begin{itemize}[leftmargin=0.1in]
		\setlength\itemsep{0.em}
		\item \textit{Weak dependence}: This claim follows from the weak dependence of the $\Delta_i$'s,  which results from the sparse (and thus locally tree-like) nature of the graph. 
		\item \textit{Gaussianity}: The right-hand side of \eqref{eq:noise} features 3 random variables, the leftmost being Gaussian and rightmost the sum of $d_i$ variables tending to an (asymptotically independent) Gaussian. It is thus reasonable that $\delta_i$ be Gaussian (so to ensure \eqref{eq:eigenvalue_eq}) yet not independent of $\Delta_i$ or $\sum_{j\in\partial_i}\delta_j$.
		\item \textit{Mean of $\delta_i$}: The symmetry of the problem at hand (equal class sizes, same affinity $c_{\rm in}$ for each class), along with the fact that the right-hand side of \eqref{eq:D-rAsigma} vanishes in its first order approximation in $d_i$, suggest that the mean of $\delta_i$ does not depend in the first order on $d_i$ but only on $\sigma_i$.  
		The amplitude of the mean then depends on $\alpha$ characterized here through $\mu_\alpha$.
		\item \textit{Variance of $\delta_i$}: The variance appears as the product of two terms: one that depends on $i$ ($\beta_i)$ and one that depends on $\alpha$. This follows from assuming that the fluctuations of $\delta_i$ follow the fluctuations of $\Delta_i$ for which the variance is similarly factorized. 
	\end{itemize} 
	
	Imposing the norm of the eigenvector $ \bm{x}_\zeta^{(2)}=\bm{\sigma} + \bm{\delta}$ to be constant with respect to $\alpha$ and the boundary condition $\mu_{\alpha_c} = 1$ (i.e., there is no information about the classes at the detectability threshold), we find the following explicit expressions for $\mu_\alpha$ and $\beta_i$.
	\begin{align*}
	1-\mu_{\alpha} &= \sqrt{\frac{c\Phi - \zeta_{\alpha}^2}{c\Phi - 1}}, \quad
	\beta_i = \frac{2}{\sqrt{d_i}}.
	\end{align*}
	Details are provided in Section~\ref{app:analysis} of the supplementary material.
	Figure~\ref{fig:simulation}-(a) supports the analysis by comparing this prediction to simulations for a synthetic network with power law degree distribution. 
	
	
	\begin{figure}[t!]
		\centering
		\includegraphics[width = \columnwidth]{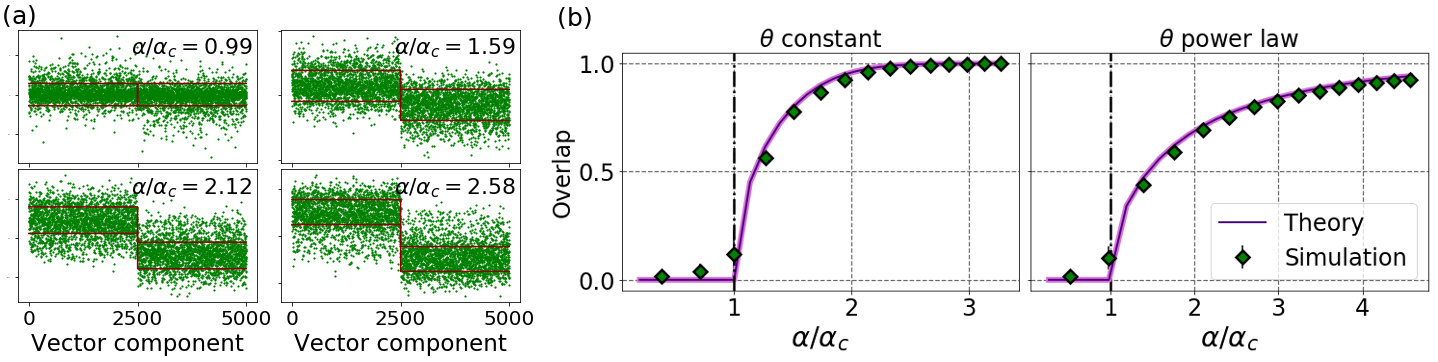}
		\caption{(a) Theoretical values of mean and variance (red line indicates $1-\mu_{\alpha} \pm 2f_{\alpha}/\sqrt{c}$) 
			vs simulation (green dots) for power-law distributed $\theta_i$'s ($\theta_i \sim Z^{-1}[\mathcal{U}(3,10)]^{4}$). 
			(b) Theoretical \eqref{eq:Ov} vs simulated overlap, averaged over $10$ realizations, for $\theta_i$ constant (left), and power-law distributed (right). 
			For both figures, $n = 5000$, $c_{\rm out} = 6$, $c_{\rm in} = 7\to 36$.}
		\label{fig:simulation}
	\end{figure}
	
	The previous line of argument provides a large dimensional approximation for the performance of spectral clustering based on the eigenvector $\bm{x}_\zeta^{(2)}$. The performance measure of interest is the \emph{overlap}, defined as ${\rm Ov} \equiv 2 \max_{\mathcal{P}_{\hat{\bm{\sigma}}}}\left[\frac{1}{n}\sum_{i = 1}^n \delta_{\sigma_i,\hat{\sigma}_i} - \frac{1}{2}\right]
	$
	where $\hat{\bm{\sigma}}$ denotes the vector of estimated labels, $\mathcal{P}_{\bm{\hat{\sigma}}}$ the set of permutations of the labels, and $\delta$ the Kronecker symbol ($\delta_{ij}=1$ if $i=j$, and $0$ otherwise). In this particularly symmetric setting \emph{only} $\hat{\sigma}_i =  {\rm sign}[(\bm{x}_\zeta^{(2)})_i]$ where $\rm sign$ is the sign function. (Remark~\ref{rem:kmeans_vs_sign} underlines the necessity not to cluster based on sign in asymmetric scenarios).
	From the expression of $\mu_\alpha$ and $\beta_i$, we find that, conditionally to $A$,
	\begin{equation}
	\label{eq:Ov}
	\mathbb{E}[{\rm Ov}] \simeq \frac{1}{n} \sum_{i = 1}^n {\rm erf}\left[\sqrt{\frac{\alpha^2d_i}{8c-2\alpha^2}\left(\frac{c\Phi - \zeta_{\alpha}^2}{c\Phi -1 }\right)}\right]
	\end{equation}
	(proof details are provided in Section~\ref{app:analysis} of the supplementary material). Figure~\ref{fig:simulation}-(b) compares the prediction of Equation~\eqref{eq:Ov} to simulations on networks with $\theta_i=1$ constant (left) or power-law distributed (right). The observed match on this $5\,000$-node synthetic network is close to perfect.
	
	\medskip
	
	As a side remark, our analysis reveals an interesting connection between $H_\zeta$ and 
	$D^{-1}A$.
	\begin{remark}[Relation to the random walk Laplacian]
		\label{rmk:rw}
		Similar to $A$, $D-A$, and $D^{-\frac12}AD^{-\frac12}$, the matrix $D^{-1}A$ is claimed \emph{inappropriate} as a spectral community detection matrix for sparse graphs. This is in fact a slight overstatement: as already observed in \cite{joseph2013impact}, as the graph under study gets sparser, $D^{-1}A$ still possesses one or possibly more informative eigenvectors, however not necessarily corresponding to dominant isolated eigenvalues (it was in particular noted that for the real network \emph{polblogs} \cite{adamic2005political} the informative eigenvector is associated to the third and not the second largest eigenvalue). This observation is easily explained in our analysis framework. Similar to our derivation for $D-\zeta A$, the average action of $D^{-1}A$ on the class vector $\bm{\sigma}$ reads
		$\mathbb{E}[[D^{-1}A\bm{\sigma}]_i|A] = \sigma_i/\zeta$
		and thus, for large $d_i$, $\bm{\sigma}$ is a close eigenvector to $D^{-1}A$, correctly predicting the existence of an informative eigenvalue also for this matrix. However, the associated eigenvalue $1/\zeta$ decays with increasing $\zeta$ and thus with harder detection tasks, hence explaining why the informative eigenvectors are associated with eigenvalues found deeper into the spectrum of $D^{-1}A$.
	\end{remark}
	
	
	
	

	
	\section{Estimating $\zeta$}
	\label{sec:estimate}
	
	While $r=\zeta$ is more appropriate a choice than $r=\sqrt{c\Phi}$, $\zeta$ is not readily accessible (as it depends on $c_{\rm in}-c_{\rm out}$), unlike $\sqrt{c\Phi}$ that is easily estimated from the $d_i$'s. To estimate $\zeta$, we elaborate on the deep relations between the Bethe Hessian $H_r$ and the non-backtracking operator $B\in \mathbb{R}^{2|\mathcal{E}|\times 2|\mathcal{E}|}$ defined, for all $(ij),(lm) \in \mathcal{E}_D$ the set of directed edges of $\mathcal{G}$, as $B_{(ij)(lm)} = \delta_{jl}(1-\delta_{im})$.  
	
	
	When $r$ is an eigenvalue of $B$, then $\det H_r = 0$ \cite{bordenave2015non,terras2010zeta}. This is convenient as $B$ only has a few isolated real eigenvalues ($B$ is non symmetric) that can send the associated isolated eigenvalues of $H_r$ to zero. This provides us with two alternative methods to estimate $\zeta$.
	
	\subsection{Exploiting the eigenvalues outside the bulk of $B$}
	
	It is proved in \cite{gulikers2016non} that, for the DC-SBM and beyond the phase transition ($\alpha>\alpha_c$), the eigenvalues $\gamma_1,\ldots,\gamma_{2m}$ of $B$, decreasingly sorted in modulus, satisfy in the large $n$ setting: $\gamma_1 \to \Phi(c_{\rm in}+c_{\rm out})/2$, $\gamma_2 \to \Phi(c_{\rm in}-c_{\rm out})/2>\sqrt{\gamma_1}$ and, for $i>2$, $\limsup_n |\gamma_i| \leq \sqrt{\gamma_1}$, almost surely.
	
	Since $\zeta=\lim_n\gamma_1/\gamma_2$, denoting $\nu_i(r)$ the eigenvalues of $H_r$ sorted in \emph{increasing} order, this result conveys the following first method to estimate $\zeta$.
	
	\begin{figure}[h!]
		\centering
		\includegraphics[width = 0.6\columnwidth]{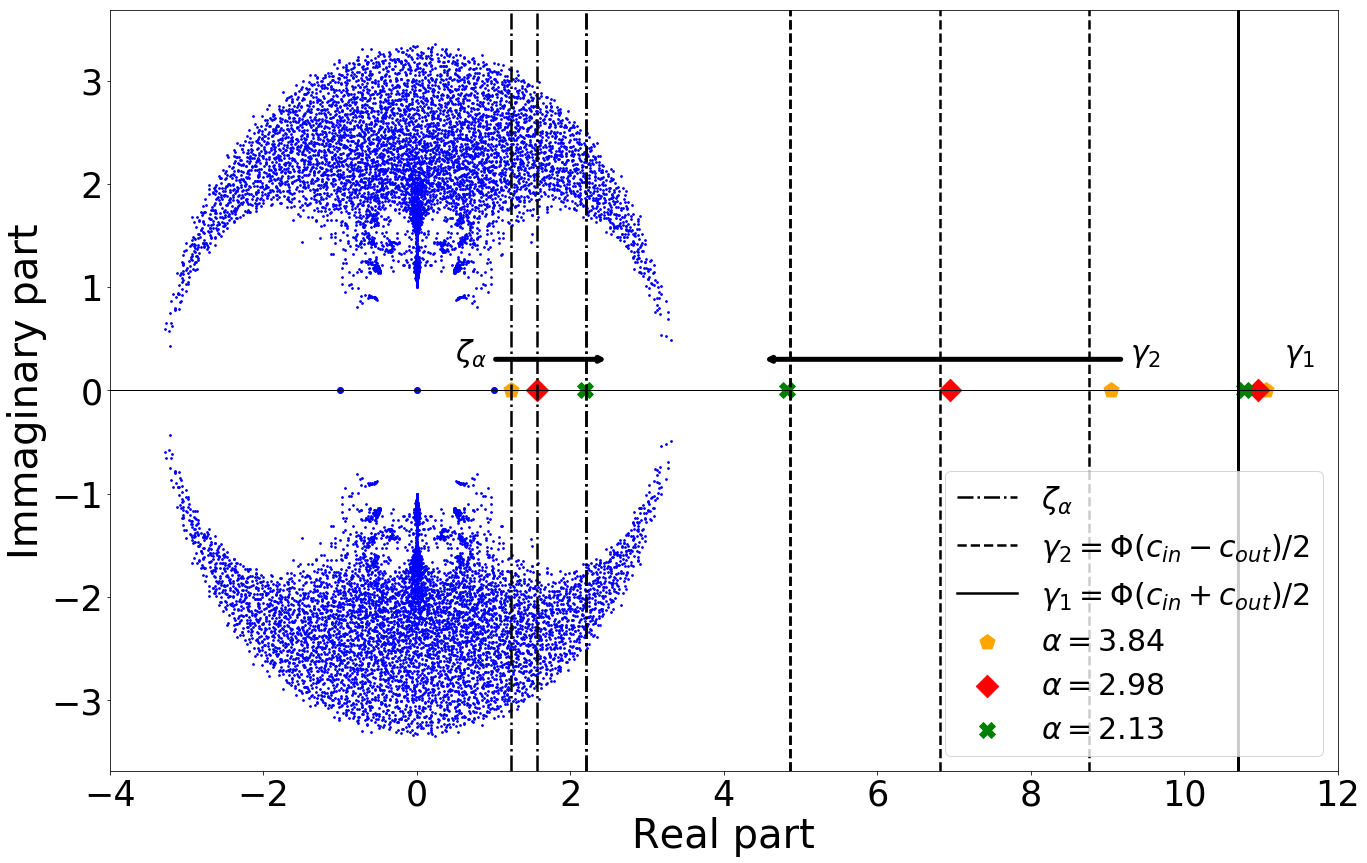}
		\caption{Superposed spectra of $B$ for 3 values of $\alpha$: $n=4000$, $c_{\rm in} = 12,11,10$ and $c_{\rm out} = 1,2,3$  ($c_{\rm in}+c_{\rm out}$ is fixed); $\bm{\theta}$ with power law distribution; all eigenvalues displayed in blue except top three dominant real displayed in colors for each $(c_{\rm in},c_{\rm out})$ pair.
		}
		\label{fig:spectrum_of_B}
	\end{figure}

	\begin{method}[First estimation of $\zeta $]
		\label{rem:est2}
		Under the previous notations $  \zeta \simeq {\gamma_1}/{\gamma_2}$.
		The eigenvalues $\gamma_1$ and $\gamma_2$ of $B$ can be estimated by a line search over $r \in(\sqrt{\rho(B)},\infty)$ on changing signs of $\nu_1(r)$ and $\nu_2(r)$ that correspond to $r = \gamma_1$ and $r = \gamma_2$, respectively.\footnote{The spectral radius of the matrix $B$, $\rho(B)$, can be estimated as $\rho(B)\simeq \sum_i d_i^2/\sum_i d_i$.}
	\end{method}
	
	\subsection{Exploiting the eigenvalues inside the bulk of $B$}
	
	The matrix $B$ can be obtained from the linearization of the belief propagation  (BP) equations (see \cite{krzakala2013spectral} for details). In particular, the linear expansion to first order of the beliefs around their fixed points yields $B \bm{w} \simeq \zeta \bm{w}$. 
	According to this argument, one expects the matrix $B$ to have a real eigenvalue equal to $\zeta$ with\footnote{This eigenvalue is visible in \cite{krzakala2013spectral, bordenave2015non} but not commented.} $\zeta \leq \sqrt{c\Phi}$. Figure~\ref{fig:spectrum_of_B} visually emphasizes this eigenvalue for three different values of $\alpha$, maintaining $c$ constant. The matrix $B$ thus has four eigenvalues inside its main bulk: $-1,0,1$ and $\zeta$. As the community detection problem gets harder, both $\zeta$ and $\gamma_2$ shift towards the edge of the bulk (from the left for the former and from the right for the latter) and then meet exactly at $\sqrt{c\Phi}$ when $\alpha=\alpha_c$. Then, for $\alpha<\alpha_c$, they reach the complex part of the bulk. 
	
	\medskip
	
	More fundamentally, simulations further suggest that the eigenvector associated with the null eigenvalue of $H_\zeta$ is precisely $\bm{x}_{\zeta}^{(2)} = \bm{\sigma}+\bm{\delta}$ studied in Section~\ref{sec:insights}. This indicates that the informative eigenvalue $\lambda_{\alpha}$ of $D-\zeta_{\alpha} A=H_{\zeta_\alpha}-(\zeta_\alpha^2-1)I_n$ in Equation~\eqref{eq:eigenvalue_eq} coincides with $-(\zeta_{\alpha}^2-1)$. It further explains why $H_{\sqrt{c\Phi}}$, initially proposed in \cite{saade2014spectral}, works well close to the detectability threshold as $\zeta \to \sqrt{c\Phi}$ when $\alpha  \to \alpha_c$. We thus expect most of the improvement of the choice $r=\zeta$ to emerge in the easier scenarios.

	Note that, as was already observed in \cite{saade2014spectral}, if $|r|>1$, then the eigenvalues of the bulk of $H_r$ are strictly positive for $|r|\neq \sqrt{c\Phi}$. As a consequence, $\bm{x}_{\zeta}^{(2)}$ is necessarily isolated when $\alpha > \alpha_c$ and so spectral clustering on $H_\zeta$ works down to the detectability threshold. To the best of our knowledge, this property is not formally proved, but we point out that it agrees with the shape of the spectrum of $B$: if the bulk of $H_r$ was negative for some $|r| > 1$, then there would be a `continuum' of real eigenvalues in $[1,\sqrt{c\Phi}]$ if $r>1$ (in the assortative case). As this is not the case, the smallest eigenvalue in the bulk of $H_r$ cannot be negative.
	
	\begin{claim}[Informative eigenvalue of $H_{\zeta_{\alpha}}$] The eigenvalue associated to the informative eigenvector of $H_{\zeta_{\alpha}}$ is equal to zero. Equivalently, the eigenvalue $\lambda_{\alpha}$ associated to the informative eigenvector of $D - \zeta_{\alpha} A$ is given by $\lambda_{\alpha} = -(\zeta_{\alpha}^2-1) = -4f^2_{\alpha} $ which vanishes for $c_{\rm out} \to 0$.
		\label{claim:1}
	\end{claim}
	
	This claim gives rise to a second method to estimate $\zeta$.
	\begin{method}[Second estimation of $\zeta$]
		\label{rem:est1}
		Under the previous notations $\nu_2(\zeta)=0$.
		The parameter $\zeta$ then corresponds to the position of change of sign of $\nu_2(r)$ in the set $r \in (1,\sqrt{\rho(B)})$.
	\end{method}
	
	\begin{figure}[t!]
		\centering
		\includegraphics[width = 0.7\columnwidth]{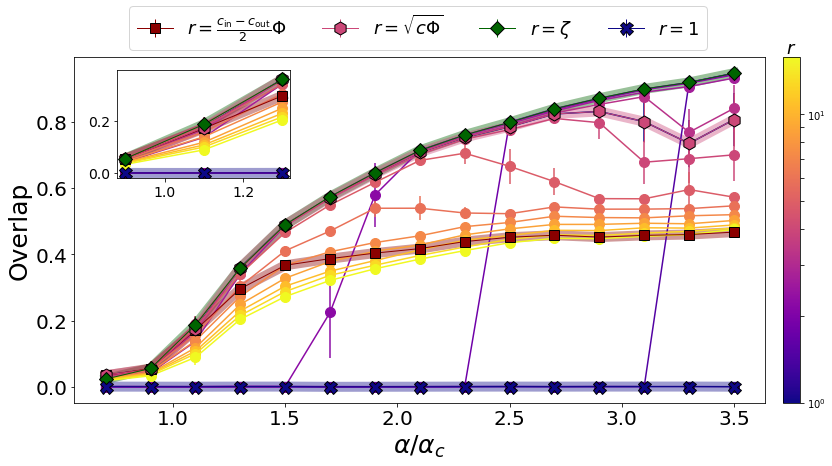}
		\caption{Overlap comparison as a function of $\alpha$, using the second smallest eigenvector of $H_r$, for different values of $r$. In color code the values of $r$ ranging from $r = 1$ (blue) to $r = c\Phi$ (yellow). The red squares indicate $r = (c_{\rm in} - c_{\rm out})\Phi/2$, that is equivalent to clustering with the matrix $B$ \cite{krzakala2013spectral}, the purple hexagons represent the Bethe-Hessian of \cite{saade2014spectral}, the green diamonds are the proposed Algorithm \ref{alg:1} and the blue crosses are the graph Laplacian. In the top left corner, a zoom of the overlap close to the transition. For these simulations, $n = 5000$, $c_{\rm in} : 15 \to 9.4$, $c_{\rm out} : 1 \to 6.6$ (while keeping $c$ fixed), $\theta_i \sim [\mathcal{U}(3,10)]^4$.}
		\label{fig:tuning}
	\end{figure}

	\section{Extension to multiple uneven-sized classes}
	\label{sec:extension}
	The analysis performed in the previous sections is resilient to heterogeneous degree distributions and can be generalized to $k$ uneven-sized classes, with last clustering step by \emph{k-means}. To this end, let $\Pi\in\RR^{k\times k}$ be diagonal with $\Pi_{ii}$ the fraction of nodes in class $i$ and assume $C\Pi \mathds{1} = c \mathds{1}$. This assumption is a standard hypothesis \cite{krzakala2013spectral,decelle2011asymptotic,bordenave2015non,zdeborova2016statistical} which ensures that the averaged node connectivity is independent of the class. For $1 \leq p \leq k$, let $(\tau_p,\bm{v}^{(p)})$ be the $p$-th largest eigenpair
	of $C\Pi$, and $\bm{u}^{(p)} \in \mathbb{R}^n$ defined as $u^{(p)}_i = v^{(p)}_{\ell_i} \: \forall \: 1 \leq i \leq n$ for $\ell_i$ the class of node $i$. The vector $\bm{u}^{(p)}$ contains plateaus with heights corresponding to the values of $\bm{v}^{(p)}$. 
	Repeating the arguments of Section~\ref{sec:choice} (see details in Section~\ref{app:more then two} of the supplementary material), we obtain $k$ choices for $r$:
	\begin{align}
	\mathbb{E}[[(D - rA)\bm{u}^{(p)}]_i] &= d_i u^{(p)}_i\left[1 - r \frac{\tau_p}{c}\right] \hspace{0.5cm}\quad \textmd{and thus}\quad
	r = \frac{c}{\tau_{p}} \equiv \zeta_p,\quad 1\leq p\leq k.
	\label{eq:opt_choice}
	\end{align}
	
	Since the largest eigenpair $(c,\mathds{1})$ of $C\Pi$  is not informative of the class structure, only the $k-1$ next largest eigenvectors $\bm{v}^{(p)}$ of $C\Pi$ are informative. The vector $\bm{u}^{(p)}$ (corresponding to the $p$-th largest eigenvalue $\tau_p$) is  in one-to-one mapping with $\bm{v}^{(p)}$ and corresponds to the $p$-th smallest value of $\zeta_p = c/\tau_p$. Considering $r = \sqrt{c\Phi}$, all the informative eigenvalues of $H_r$ are negative \cite{saade2014spectral}. By decreasing $r$ they progressively become positive: for $r = \zeta_k$ (the largest among $\zeta_p$) the $k$-th smallest eigenvalue is the first to hit zero. By further decreasing $r$, all the informative eigenvalues follow, until $r = \zeta_1 = 1$ for which the smallest eigenvalue is null. We conclude that $\bm{u}^{(p)}$  is  associated with the $p$-th smallest eigenvector $\bm{x}_{\zeta_p}^{(p)}$ of $H_{\zeta_p}$.

	Method~\ref{rem:est2} and Method~\ref{rem:est1} both generalize to this scenario. In particular the outer eigenvalues of $B$ converge as $\gamma_p \to \tau_p\Phi$ and the linearization of BP retrieves the fixed points as $\zeta_p = c/\tau_p$.
	
	For $k > 2$, the value $r = \sqrt{c\Phi}$ still plays an important role. It was chosen in \cite{saade2014spectral} because, asymptotically, for this value of $r$ \emph{only} the informative eigenvalues of $H_{\sqrt{c\Phi}}$ are negative. The number of classes is then directly obtained from counting the number of negative eigenvalues of $H_{\sqrt{c\Phi}}$. The relation between $H_r$ and $B$ further guarantees that the number of isolated eigenvalues of $B$ (hence of $H_r$) is asymptotically equal to the number of detectable classes. 

	\begin{remark}[On k-means versus signed-based clustering]
		\label{rem:kmeans_vs_sign}
		Under a symmetric $2$-class of even size setting, the classification of the entries of the informative eigenvector of $H_r$ can be performed based on their signs. This sign-based method first does not generalize to more than two or uneven sized classes, where k-means or expectation-maximization based clustering is required. But it also hinders the fact that the eigenvector entries may be quite concentrated around zero (close to $0^+$ or $0^-$ according to the class) and thus \emph{not clustered}, a situation where k-means has no discriminative power.
		
		Simulations (and reported results in \cite{saade2014spectral} based on signs rather than k-means) suggest that the informative eigenvector of $H_{\sqrt{c\Phi}}$ precisely suffers this condition. We have demonstrated here instead that the informative eigenvector of $H_\zeta$ has the convenient feature of being genuinely clustered.
	\end{remark}
	
	\begin{figure}[t]
		\centering
		\includegraphics[width = \columnwidth]{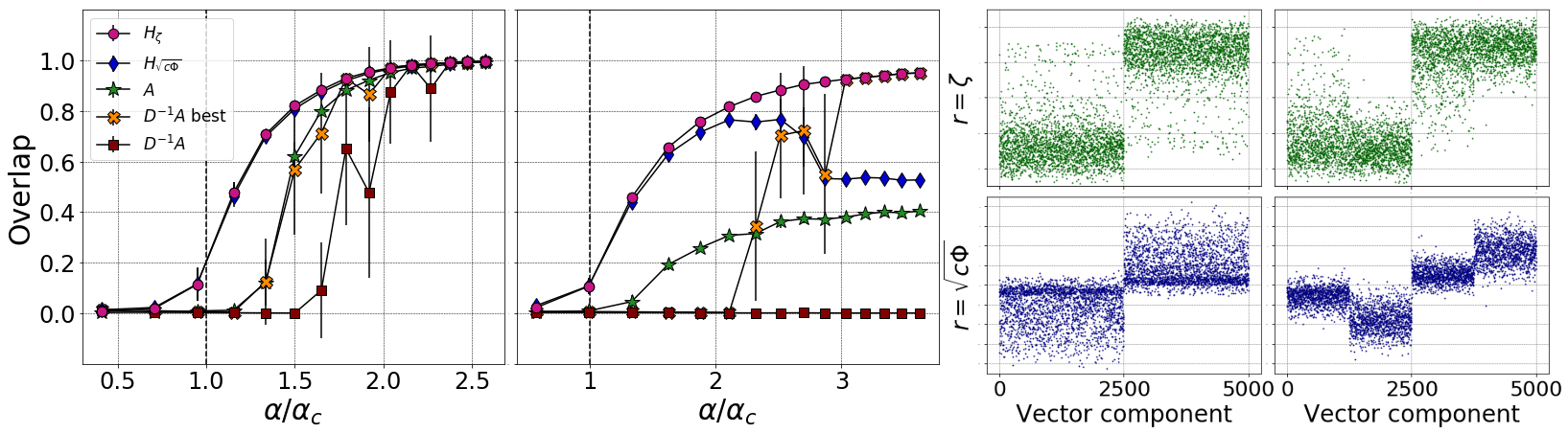}
		\caption{(a) Comparison of spectral clustering for $\theta_i=1$ (left) and with power law distribution $\theta_i \sim Z^{-1}[\mathcal{U}(3,10)]^4$. ``$D^{-1}A$ best'' indicates spectral clustering on the best (among the first 25) eigenvector of $D^{-1}A$. Here, $n = 5000$, $c_{\rm out} = 1$, $c_{\rm in} = 2 \to 16$. Averaged over $10$ samples. 
			The error bars indicate one standard deviation. 
			(b) 
			$\bm{x}_{\zeta}^{(2)}$ (top) and $\bm{x}_{\sqrt{c\Phi}}^{(2)}$ (bottom) for power law distributed $\theta_i$ (left) and for $\theta_i=\theta_0$, $i\leq n/4$ and $n/2 \leq i \leq 3n/4$, and $\theta_i=4\theta_0$ otherwise (right).}
		\label{fig:all_methods}
	\end{figure}

	\section{Experimental validation}
	\label{sec:validation}
	
	Our results can be summarized by Algorithm \ref{alg:1}, where we recall that $\nu_p(r)$ is the $p$-th smallest eigenvalue of $H_r$ and where $\bm{x}_r^{(p)}$ indicates the corresponding eigenvector.

	Figure~\ref{fig:tuning} depicts the overlap, as a function of $\alpha$, of the output of a two-class $k$-means on the informative eigenvector of $H_r$, for different values of $r$, ranging from $1$ to $c\Phi$. When $\alpha$ is large enough, small values of $r$ lead to better partitions than large values of $r$ that are more affected by degree heterogeneity. However, for $r$ small, the informative eigenvector is not necessarily corresponding to the second smallest eigenvalue, leading to a meaningless partition. On the contrary, larger values of $r$ show isolated eigenvectors also in the "hard regime". We recall that $r = \zeta$ is an $\alpha$-dependent parameter: for $\alpha \to \alpha_c$, $\zeta$ is "large  enough" so that the informative eigenvalue is isolated, while for $\alpha \to \sqrt{2c_{\rm in}}$  it is "small enough" to give good partitions. Also the value of $r = (c_{\rm in} - c_{\rm out})\Phi/2$ is $\alpha$-dependent and it corresponds to clustering with $B$ as indicated in \cite{krzakala2013spectral}. While it gives good partitions very close to the transition, this choice of $r$ seems largely sub-optimal for easier tasks.

	Figure~\ref{fig:all_methods}-(a) compares the overlaps obtained with Algorithm~\ref{alg:1} versus related spectral clustering methods based on $H_{\sqrt{c\Phi}}$, $D^{-1}A$ and $A$. Accordingly with Remark~\ref{rem:kmeans_vs_sign}, \emph{k-means} clustering (rather than sign-based) on the informative eigenvectors is systematically performed.
	For $\theta_i=1$, the left display recovers the results of \cite{saade2014spectral}, evidencing a strong advantage for $H_r$ versus Laplacian methods. Since the degrees are similar, both $r=\sqrt{c\Phi}$ and $r=\zeta$ induce similar $H_r$ performances. The improvement provided by $H_\zeta$ arises in the right display for power-law distributed $\theta_i$, with most of the gain appearing away from the detection threshold.
	On both displays is also depicted the performance of $D^{-1}A$ based on its second largest eigenvector and on an oracle choice of the informative eigenvector with maximal overlap. These curves confirm Remark~\ref{rmk:rw} on the non-dominant position of the informative eigenvector of $D^{-1}A$ in hard tasks.\footnote{The low performance of $D^{-1}A$, even in an oracle setting, can be attributed to the high density of eigenvalues in the bulk of the spectrum which induces a ``dispersion'' of the informative eigenvectors to the eigenvectors associated to neighboring eigenvalues. The class information is thus ``spread'' across several eigenvectors.}
	Figure~\ref{fig:all_methods}-(b) depicts the informative eigenvectors of $H_{\sqrt{c\Phi}}$ and $H_\zeta$, demonstrating the negative impact of $\theta_i$ on $H_{\sqrt{c\Phi}}$, in stark contrast with the resilience of $H_\zeta$.
	
	Table~\ref{tab:real} next provides a comparison of the algorithm performances on real networks, both labelled and unlabelled, confirming the overall superiority of Algorithm~\ref{alg:1}, quite unlike $H_{\sqrt{c\Phi}}$ which fails on several examples.\footnote{In Table~\ref{tab:real}, the modularity is defined as $\mathcal{M} = \frac{1}{2|\mathcal{E}|} \sum_{i,j =1}^n\left(A_{ij}-\frac{d_id_j}{2|\mathcal{E}|}\right)\delta(\hat{\ell}_i,\hat{\ell}_j)$, see e.g., \cite{newman2004finding,Newman-2006}.} 
	
	\begin{algorithm}[t!]
		\begin{algorithmic}[1]
			\State \textbf{Input} : adjacency matrix of undirected graph $\mathcal{G}$
			\State Detect the number of classes:  $\hat{k} \leftarrow |\{i,~\nu_i(\sqrt{c\Phi}) < 0\}|$.
			\For{$2 \leq p \leq \hat{k}$}
			\State  $\zeta_p \leftarrow r$ such that $\nu_p(r) = 0$
			\State  $X_p \leftarrow \bm{x}_{\zeta_p}^{(p)}$
			\EndFor
			\State Estimate community labels $\hat{\bm{\ell}}$ as output of $\hat{k}$-class \emph{k-means} on the rows of $X = [X_2, \ldots, X_{\hat{k}}]$.
			\Return Estimated number $\hat{k}$ of communities and label vector $\hat{\bm{\ell}}$.
			\caption{Improved Bethe-Hessian Community Detection}
			\label{alg:1}
		\end{algorithmic}
	\end{algorithm}
	
	
	\begin{table}[h!]
		\small
		\hspace*{-.5cm}
		\begin{tabular}{|c||c|c||c|c|c|||c||c|c||c|c|c|c|}
			\hline
			L & $n$ & $k$ & Alg.\ref{alg:1} 
			&  $H_{\sqrt{c\Phi}}$ & $A$ & U & $n$ & $k$ & Alg.\ref{alg:1}
			& $H_{\sqrt{c\Phi}}$ & $A$ \\
			\hline
			\hline
			Karate \cite{zachary1977information}  & 34 & 2 & $\bm{1.00}$ & $\bm{1.00}$ & $\bm{1.00}$ & Mail  & 1133 & 21 & $\bm{0.50}$ & $0.40$ & $0.32$ \\
			\hline
			Dolphins \cite{lusseau2003bottlenose}  & 62 & 2 & $\bm{0.97}$  & $0.87$ & $0.65$ & Facebook  & 4039 & 65 & $\bm{0.77}$ & $0.48$ & $0.38$ \\
			\hline
			Polbooks \cite{orgnet} & 105 & 3 & $\bm{0.77}$  & $0.74$ & $0.57$ & Power grid  & 4941 & 53 & $\bm{0.92}$ & $0.61$ & $0.31$ \\
			\hline
			Football \cite{girvan2002community} & 115 & 12 & $\bm{0.92}$ & $\bm{0.92}$ & $\bm{0.92}$  & Nutella  & 6301 & 5 & $\bm{0.34}$ & $0.15$ & $0.14$\\
			\hline
			Polblogs \cite{adamic2005political}  & 1221 & 2  & $\bm{0.91}$ & 0.32 & 0.26 & Wikipedia  & 7115 & 21 & $\bm{0.21}$ & 0.$18$ & $0.15$ \\
			\hline
		\end{tabular}
		\vspace{0.15cm}
		\caption{Performance comparison on real networks. Labelled datasets with $k$ known and overlap comparison:  (left). Unlabelled networks \cite{snapnets} with $k$ estimated and modularity comparison. Only assortative features are kept into account.}
		\label{tab:real}
	\end{table}
	\vspace{-0.5cm}

	

	\section{Concluding Remarks}
	\label{sec:conclusion}

	
	
	Beyond the demonstration of superiority of $H_\zeta$ to $H_{\sqrt{c\Phi}}$, originally proposed in \cite{saade2014spectral}, the article provides a consistent understanding of the natural limitations and strengths of the wide class of spectral clustering methods involving combinations of $A$ and $D$.
	
	Yet, other methods, the performances of which cannot always be compared on even grounds, have been proposed in the literature that marginally relate to the present study. This is notably the case of \cite{qin2013regularized} which performs spectral clustering on $L_\tau=(D+\tau I_n)^{-\frac12}A(D+\tau I_n)^{-\frac12}$ (with a proposed choice $\tau=c$) which aims at neutralizing the deleterious effects of small $d_i$. Although evidently affecting the spectrum (and thus the informative structure) of $A$ by the non-linear normalization, simulations on $L_\tau$ suggest competitive performances to $H_\zeta$ in almost all studied examples. A systematic analysis of this and similarly proposed methods in the literature is clearly called for.
	
	Despite its demonstrated significant performance improvement, Algorithm~\ref{alg:1} suffers from a slightly larger computational cost than most competing methods ($O(nk^3)$ instead of the usual $O(nk^2)$ complexity in the case of sparse graph) due to the successive estimations of $\zeta$. 
	We are currently working on improving this computation time. 
	
	From a theoretical standpoint, the request for $c\gg 1$ is still inappropriate to many practical networks. A first consequence of smaller values for $c$ is the loss of Gaussianity of the eigenvector entries as already evidenced in Figures~\ref{fig:simulation} and \ref{fig:all_methods} where Gaussianity is clearly lost in the easiest tasks in profit of a ``one-sided'' distribution. This suggests further improvement of our analysis framework along with the development of algorithms more appropriate than k-means to handle the last clustering step.
	
	\subsection*{Acknowledgments}
	This work is supported by the ANR Project RMT4GRAPH (ANR-14-CE28-0006), the IDEX GSTATS Chair at University Grenoble Alpes and by CNRS PEPS I3A (Project RW4SPEC). The authors thank Jean-Louis Barrat for fruitful discussions.
	

\begin{thebibliography}{10}
	
	\bibitem{fortunato2010community}
	Santo Fortunato.
	\newblock Community detection in graphs.
	\newblock {\em Physics reports}, 486(3-5):75--174, 2010.
	
	\bibitem{mossel2014belief}
	Elchanan Mossel, Joe Neeman, and Allan Sly.
	\newblock Belief propagation, robust reconstruction and optimal recovery of
	block models.
	\newblock In {\em Conference on Learning Theory}, pages 356--370, 2014.
	
	\bibitem{rohe2011spectral}
	Karl Rohe, Sourav Chatterjee, Bin Yu, et~al.
	\newblock Spectral clustering and the high-dimensional stochastic blockmodel.
	\newblock {\em The Annals of Statistics}, 39(4):1878--1915, 2011.
	
	\bibitem{von2007tutorial}
	Ulrike Von~Luxburg.
	\newblock A tutorial on spectral clustering.
	\newblock {\em Statistics and computing}, 17(4):395--416, 2007.
	
	\bibitem{gulikers2017spectral}
	Lennart Gulikers, Marc Lelarge, and Laurent Massouli{\'e}.
	\newblock A spectral method for community detection in moderately sparse
	degree-corrected stochastic block models.
	\newblock {\em Advances in Applied Probability}, 49(3):686--721, 2017.
	
	\bibitem{lei_consistency_2015}
	J.~Lei and A.~Rinaldo.
	\newblock Consistency of spectral clustering in stochastic block models.
	\newblock {\em The Annals of Statistics}, 43(1):215--237, 2015.
	
	\bibitem{nadakuditi2012graph}
	Raj~Rao Nadakuditi and Mark~EJ Newman.
	\newblock Graph spectra and the detectability of community structure in
	networks.
	\newblock {\em Physical review letters}, 108(18):188701, 2012.
	
	\bibitem{ali2016random}
	Hafiz {Tiomoko Ali} and Romain Couillet.
	\newblock Random matrix improved community detection in heterogeneous networks.
	\newblock In {\em Signals, Systems and Computers, 2016 50th Asilomar Conference
		on}, pages 1385--1389. IEEE, 2016.
	
	\bibitem{saade2014spectral}
	Alaa Saade, Florent Krzakala, and Lenka Zdeborov{\'a}.
	\newblock Spectral clustering of graphs with the bethe hessian.
	\newblock In {\em Advances in Neural Information Processing Systems}, pages
	406--414, 2014.
	
	\bibitem{krzakala2013spectral}
	Florent Krzakala, Cristopher Moore, Elchanan Mossel, Joe Neeman, Allan Sly,
	Lenka Zdeborov{\'a}, and Pan Zhang.
	\newblock Spectral redemption in clustering sparse networks.
	\newblock {\em Proceedings of the National Academy of Sciences},
	110(52):20935--20940, 2013.
	
	\bibitem{bordenave2015non}
	Charles Bordenave, Marc Lelarge, and Laurent Massouli{\'e}.
	\newblock Non-backtracking spectrum of random graphs: community detection and
	non-regular ramanujan graphs.
	\newblock In {\em Foundations of Computer Science (FOCS), 2015 IEEE 56th Annual
		Symposium on}, pages 1347--1357. IEEE, 2015.
	
	\bibitem{massoulie2014community}
	Laurent Massouli{\'e}.
	\newblock Community detection thresholds and the weak ramanujan property.
	\newblock In {\em Proceedings of the forty-sixth annual ACM symposium on Theory
		of computing}, pages 694--703. ACM, 2014.
	
	\bibitem{mossel2015reconstruction}
	Elchanan Mossel, Joe Neeman, and Allan Sly.
	\newblock Reconstruction and estimation in the planted partition model.
	\newblock {\em Probability Theory and Related Fields}, 162(3-4):431--461, 2015.
	
	\bibitem{barabasi1999emergence}
	Albert-L{\'a}szl{\'o} Barab{\'a}si and R{\'e}ka Albert.
	\newblock Emergence of scaling in random networks.
	\newblock {\em science}, 286(5439):509--512, 1999.
	
	\bibitem{gulikers2016non}
	Lennart Gulikers, Marc Lelarge, and Laurent Massouli{\'e}.
	\newblock {Non-Backtracking Spectrum of Degree-Corrected Stochastic Block
		Models}.
	\newblock In Christos~H. Papadimitriou, editor, {\em 8th Innovations in
		Theoretical Computer Science Conference (ITCS 2017)}, volume~67 of {\em
		Leibniz International Proceedings in Informatics (LIPIcs)}, pages
	44:1--44:27, Dagstuhl, Germany, 2017. Schloss Dagstuhl--Leibniz-Zentrum fuer
	Informatik.
	
	\bibitem{gulikers2015impossibility}
	Lennart Gulikers, Marc Lelarge, Laurent Massouli{\'e}, et~al.
	\newblock An impossibility result for reconstruction in the degree-corrected
	stochastic block model.
	\newblock {\em The Annals of Applied Probability}, 28(5):3002--3027, 2018.
	
	\bibitem{karrer2011stochastic}
	Brian Karrer and Mark~EJ Newman.
	\newblock Stochastic blockmodels and community structure in networks.
	\newblock {\em Physical review E}, 83(1):016107, 2011.
	
	\bibitem{qin2013regularized}
	Tai Qin and Karl Rohe.
	\newblock Regularized spectral clustering under the degree-corrected stochastic
	blockmodel.
	\newblock In {\em Advances in Neural Information Processing Systems}, pages
	3120--3128, 2013.
	
	\bibitem{le2017concentration}
	Can~M Le, Elizaveta Levina, and Roman Vershynin.
	\newblock Concentration and regularization of random graphs.
	\newblock {\em Random Structures \& Algorithms}, 51(3):538--561, 2017.
	
	\bibitem{joseph2013impact}
	Antony Joseph and Bin Yu.
	\newblock Impact of regularization on spectral clustering.
	\newblock {\em arXiv preprint arXiv:1312.1733}, 2013.
	
	\bibitem{dembo2010gibbs}
	Amir Dembo, Andrea Montanari, et~al.
	\newblock Gibbs measures and phase transitions on sparse random graphs.
	\newblock {\em Brazilian Journal of Probability and Statistics},
	24(2):137--211, 2010.
	
	\bibitem{decelle2011asymptotic}
	Aurelien Decelle, Florent Krzakala, Cristopher Moore, and Lenka Zdeborov{\'a}.
	\newblock Asymptotic analysis of the stochastic block model for modular
	networks and its algorithmic applications.
	\newblock {\em Physical Review E}, 84(6):066106, 2011.
	
	\bibitem{adamic2005political}
	Lada~A Adamic and Natalie Glance.
	\newblock The political blogosphere and the 2004 us election: divided they
	blog.
	\newblock In {\em Proceedings of the 3rd international workshop on Link
		discovery}, pages 36--43. ACM, 2005.
	
	\bibitem{terras2010zeta}
	Audrey Terras.
	\newblock {\em Zeta functions of graphs: a stroll through the garden}, volume
	128.
	\newblock Cambridge University Press, 2010.
	
	\bibitem{zdeborova2016statistical}
	Lenka Zdeborov{\'a} and Florent Krzakala.
	\newblock Statistical physics of inference: Thresholds and algorithms.
	\newblock {\em Advances in Physics}, 65(5):453--552, 2016.
	
	\bibitem{newman2004finding}
	Mark~EJ Newman and Michelle Girvan.
	\newblock Finding and evaluating community structure in networks.
	\newblock {\em Physical review E}, 69(2):026113, 2004.
	
	\bibitem{Newman-2006}
	M.~E.~J. Newman.
	\newblock Modularity and community structure in networks.
	\newblock {\em Proceedings of the National Academy of Sciences},
	103:8577--8582, 2006.
	
	\bibitem{zachary1977information}
	Wayne~W Zachary.
	\newblock An information flow model for conflict and fission in small groups.
	\newblock {\em Journal of anthropological research}, 33(4):452--473, 1977.
	
	\bibitem{lusseau2003bottlenose}
	David Lusseau, Karsten Schneider, Oliver~J Boisseau, Patti Haase, Elisabeth
	Slooten, and Steve~M Dawson.
	\newblock The bottlenose dolphin community of doubtful sound features a large
	proportion of long-lasting associations.
	\newblock {\em Behavioral Ecology and Sociobiology}, 54(4):396--405, 2003.
	
	\bibitem{orgnet}
	http://www.orgnet.com/.
	
	\bibitem{girvan2002community}
	Michelle Girvan and Mark~EJ Newman.
	\newblock Community structure in social and biological networks.
	\newblock {\em Proceedings of the national academy of sciences},
	99(12):7821--7826, 2002.
	
	\bibitem{snapnets}
	Jure Leskovec and Andrej Krevl.
	\newblock {SNAP Datasets}: {Stanford} large network dataset collection.
	\newblock \url{http://snap.stanford.edu/data}, June 2014.
	
\end{thebibliography}

	\bibliographystyle{unsrt}

	 \renewcommand{\thesubsection}{\Alph{subsection}}
\def\theequation{\thesection.\arabic{equation}}
\numberwithin{equation}{subsection}

\section*{Supplementary material}
	\subsection{\label{app:map to ising}Mapping to Ising}
	
	As introduced in \cite{decelle2011asymptotic} for the SBM ($\bm{\theta} = \mathds{1}_n$), the probability to realize a graph under the sparse DC-SBM hypothesis reads:
	\begin{align*}
	\mathbb{P}(A|\bm{\sigma},\bm{\theta}) &= \prod_{i, j<i} \left(\theta_i\theta_j \frac{C_{\sigma_i,\sigma_j}}{n}\right)^{A_{ij}}  \left(1-\theta_i\theta_j \frac{C_{\sigma_i,\sigma_j}}{n}\right)^{1-A_{ij}}
	= \prod_{i,j<i} \left(\theta_i\theta_j \frac{C_{\sigma_i,\sigma_j}}{n}\right)^{A_{ij}} + o\left(\frac{1}{n}\right)   \\
	&= \prod_{(ij) \in \mathcal{E}} \theta_i\theta_j \frac{C_{\sigma_i,\sigma_j}}{n} + o\left(\frac{1}{n}\right) .
	\end{align*}
	
	By making use of the Bayes theorem we can map the probability distribution of the labels to a physical analogue of spins interacting on the graph.
	\begin{align*}
	\mathbb{P}(\bm{\sigma}|A) &= \int d\bm{\theta} \: \mathbb{P}(\bm{\sigma},\bm{\theta}|A)
	= \int d\bm{\theta} \: \mathbb{P}(A|\bm{\sigma},\bm{\theta}) \frac{\mathbb{P}(\bm{\sigma})\mathbb{P}(\bm{\theta})}{\mathbb{P}(A)} \\ 
	&\underset{n \to \infty}{\sim} \frac{1}{\mathbb{P}(A)2^n}  \prod_{(ij) \in \mathcal{E}} \frac{C_{\sigma_i,\sigma_j}}{n} \int d\bm{\theta} \: \mathbb{P}(\bm{\theta}) \theta_i \theta_j 
	=\frac{1}{Z}\prod_{(ij)\in \mathcal{E}} C_{\sigma_i,\sigma_j}
	= \frac{1}{Z} e^{-\tilde{\mathcal{H}}(\bm{\sigma})}
	\end{align*}
	where we recovered the Boltzmann distribution with dimensionless Hamiltonian given by
	\begin{equation}
	\mathcal{\tilde{H}}(\bm{\sigma}) = -\sum_{(ij) \in \mathcal{E}} \log\left[C_{\sigma_i,\sigma_j}\right] \equiv -\sum_{(ij) \in \mathcal{E}} {\rm ath}\left(\frac{1}{r}\right) \sigma_i\sigma_j + const
	\label{eq:ising}
	\end{equation}
	where $const$ is a constant that will be absorbed in the normalization factor. This last step gives rise to an Ising Hamiltonian. The following system of equations must then hold for some $r$:
	\begin{subeqnarray}
		\log[c_{\rm in}] &=& {\rm ath}\left(\frac{1}{r}\right) + const. \\
		\log[c_{\rm out}] &=& -{\rm ath}\left(\frac{1}{r}\right) + const.
	\end{subeqnarray}
	It is easy to check that $r = \zeta$ is the solution to this system of equations. From this result, one can then follow the derivation of the Bethe-Hessian matrix proposed in \cite{saade2014spectral}. 
	
	It has to be remarked that to obtain Equation~\eqref{eq:ising} we neglected terms coming from non-nearest neighbours, in the limit  for $n \to \infty$. The mapping is therefore not exact, but it still constitutes a useful tool to analyze and understand the problem.
	
	Further note that, for disassortative networks, $c_{\rm in} < c_{\rm out}$ and thus $\zeta < 0$ as commented in Remark~\ref{rem:dis} in the main article. This would correspond to an anti-ferromagnetic interaction between the spins, in complete agreement  with the mapping provided.

	\subsection{Mean and variance of the eigenvector}
	\label{app:analysis}
	We need to identify the terms $\beta_i$ and $\mu_\alpha$ introduced in Assumption~\ref{ass:delta_i} to track the behavior of $\bm{\delta}$ and thus of the eigenvector $\bm{\sigma}+\bm{\delta}$. A first constraint on $\bm\delta$ follows from imposing the normalization of the eigenvector which, in the trivial limit equals $\bm\sigma$, the norm of which is $\sqrt{n}$. As such,
	\begin{equation}
	\Vert(1-\mu_{\alpha})\bm{\sigma} + f_{\alpha}\bm{\beta}\odot \bm{N}\Vert^2 = n
	\label{eq:norm}
	\end{equation}
	where $\bm{\beta}=(\beta_i)_{i=1}^n$, and $\bm{N}$ is a vector of zero mean and unit variance Gaussian random variables. 
	Denoting $n\tilde{\beta}^2 \equiv \Vert\bm{\beta} \odot \bm{N}\Vert^2$ and observing that $\tilde{\beta} = O(\beta_i)$ -- \emph{i.e.} they have the same scaling with respect to $c$ --, we can rewrite this equation under the form:
	\begin{equation}
	\label{eq:norm2}
	(1-\mu_{\alpha})^2 + f^2_{\alpha}\tilde{\beta}^2 = 1.
	\end{equation}
	This provides a first relation between $\mu_\alpha$ and $\tilde{\beta}$. To obtain our next equations, we now explore boundary conditions on the model parameters in the limit of trivial clustering and at the phase transition where clustering becomes impossible.
	
	It is established in \cite{gulikers2015impossibility} that there exists a critical value $\alpha_c\equiv 2/\sqrt{\Phi}$ for $\alpha$
	below which community detection is (asymptotically) impossible. In particular, for $\alpha= \alpha_c$, the eigenvector $\bm{\sigma}+\bm{\delta}$ does not contain any information about the classes and thus $\mu_{\alpha_c} = 1$. From Equation~\eqref{eq:f_alpha}, we then find that $
	f_{\alpha_c} = \sqrt{c\Phi-1}/2 $.
	Also, from \eqref{eq:norm2}, we get $\tilde{\beta} = 1/f_{\alpha_c}$. Updating \eqref{eq:norm2}, we now have an explicit expression for $\mu_{\alpha}$ for all $\alpha$. Recalling that $4f^2_{\alpha} = \zeta_{\alpha} ^2 -1$ (from \eqref{eq:def_zeta} and \eqref{eq:f_alpha}) then gives
	\begin{equation}
	1- \mu_{\alpha} = \sqrt{\frac{c\Phi-\zeta_{\alpha} ^2}{c\Phi-1}} .
	\end{equation}
	
	Getting back to \eqref{eq:eigenvalue_eq} and \eqref{eq:noise}, it now remains to estimate $\beta_i$, which we shall perform in the limit $\alpha \to \sqrt{2c_{\rm in}}$ of trivial clustering. To this end, combining both equations, we have
	\begin{align*}
	2f_{\alpha}(1-\mu_{\alpha})\sqrt{d_i}\tilde{N_i}- \zeta_{\alpha}  \sum_{j \in \partial_i}f_{\alpha}\beta_j N_j + d_if_{\alpha}\beta_i N_i = \lambda_{\alpha}[(1-\mu_{\alpha})\sigma_i + f_{\alpha}\beta_i N_i]
	\end{align*}
	for $\tilde{N}_1, N_1,\ldots, \tilde{N}_n,N_n$ all (non necessarily independent) standard normal random variables. The second left-hand side term is proportional to $\sqrt{d_i}$ (and thus of order $O(\sqrt{c})$) as per the weak independence assumption of the $N_k$'s (Assumption~\ref{ass:delta_i}). Dividing both sides by $f_{\alpha}\sqrt{d_i}$ to equate terms of order $O(1)$, the right-hand side now scales as $\lambda_{\alpha}/(f_{\alpha}\sqrt{d_i})$. As noted in Remark~\ref{rem:consistency}, in the trivial clustering limit where $\alpha\to\sqrt{2c_{\rm in}}$, $\lambda_{\alpha} \to 0$, but it is not clear whether the right-hand side (after division by $f_{\alpha}\sqrt{d_i}$) vanishes; we now investigate this term in detail.
	One may at first observe that, if $c_{\rm out} = \epsilon c_{\rm in}$ for $\epsilon \ll 1$, since $c$ typically scales like $d_i$, we obtain that $f_{\alpha}\sqrt{d_i} = \sqrt{\epsilon c_{\rm in}/2} + O(\epsilon)$. 
	Hence, if $c_{\rm in} \gtrapprox \epsilon^{-1}$, the right-hand side vanishes. But imposing this growth condition is in fact not even necessary. If $\lambda_{\alpha} \propto f^\eta_{\alpha}$ for some $\eta > 1$, we directly obtain a vanishing right-hand side term; in Section \ref{sec:estimate} we argued that $\eta = 2$ (see Claim~\ref{claim:1}). \\
	Denoting $\sum_{j \in \partial_i} \beta_j N_j \equiv \langle\beta\rangle N \sqrt{d_i}$ for some $\langle\beta\rangle>0$, 
	we may then rewrite
	\begin{equation}
	\label{eq:advanced_eq}
	2(1-\mu_{\alpha})\tilde{N}_i - \zeta_{\alpha}  \langle\beta\rangle N + \sqrt{d_i} \beta_i N_i \to 0
	\end{equation}
	in the limit $\alpha\to\sqrt{2c_{\rm in}}$. Besides, $\mu_{\alpha} \to 0$ while $\zeta_{\alpha}  \to 1$. We already argued that $\beta_i$ (and thus $\langle\beta\rangle$), which is of the order of $\tilde{\beta}$, scales as $1/f_{\alpha_c}=O(c^{-1/2})$. Thus, in the  limit of large degrees, the second term in \eqref{eq:advanced_eq} is negligible and the third of order $O(1)$. Equating the large degree limiting variances of the resulting equation finally gives
	\begin{align*}
	\beta_i = \frac2{\sqrt{d_i}}.
	\end{align*}
	We now have the mean and the variance of each vector component and we can estimate the expression of the overlap.
	Considering a node with $\sigma_i = 1$ without loss of generality, in the large $c$ limit, we have the approximate classification error for node $i$:
	\begin{align*}
	\mathbb{P}^{i}_{\rm err} &\simeq \frac{1}{\sqrt{2\pi [f_{\alpha}\beta_i]^2}}\int_{(1-\mu_{\alpha})}^{\infty} e^{-{x^2}/(2[f_{\alpha}\beta_i]^2)}dx
	= \frac{1}{2}\left[1-{\rm erf}\left(\frac{1}{\sqrt{2}[f_{\alpha}\beta_i]}(1-\mu_{\alpha})\right)\right].
	\end{align*}
	From this, the expression of the overlap follows.
	
	\subsection{Extension to more than two classes}
	\label{app:more then two}
	In order to generalize the argument carried on for two classes, first we look into the following quantity
	\begin{align*}
	\mathbb{P}(\ell_i|\ell_j,A_{ij}=1) &= \frac{\mathbb{P}(\ell_i,\ell_j|A_{ij}=1)}{\mathbb{P}(\ell_j|A_{ij}=1)}
	=\frac{\iint d\theta_i d\theta_j \mathbb{P}(\ell_i,\ell_j,\theta_i\theta_j|A_{ij}=1)}{\mathbb{P}(\ell_j)} \\
	&= \frac{\iint d\theta_id\theta_j 	\mathbb{P}(A_{ij}=1|\theta_i,\theta_j,\ell_i,\ell_j)\mathbb{P}(\ell_i)\mathbb{P}(\ell_j)\mathbb{P}(\theta_i)\mathbb{P}(\theta_j)}{Z \pi_{\ell_j}}\\
	&= \frac{\pi_{\ell_i}C_{\ell_i,\ell_j}}{c} =  \frac{(\Pi C)_{\ell_i,\ell_j}}{c} =  \frac{(C \Pi)_{\ell_j,\ell_i}}{c }
	\end{align*}
	
	By repeating the same argument on the average behavior of the adjacency matrix we obtain:
	\begin{align*}
	\langle(A\bm{u}^{(p)})_i\rangle &= \sum_{j \in \partial(i)} \langle u_j^{(p)}\rangle = \sum_{j \in \partial(i)} \langle v_{\ell_j}^{(p)} \rangle  
	= d_i \sum_{\ell_j} \mathbb{P}(\ell_j|\ell_i,A_{ij}=1) v_{\ell_j}^{(p)}   \\
	&= \frac{d_i}{c} \sum_{\ell_j}(C\Pi)_{\ell_i,\ell_j}v_{\ell_j}^{(p)} 
	= \frac{d_i}{c} (C\Pi v^{(p)})_{\ell_i} = \frac{d_i}{c} \tau_{p} v_{\ell_i}^{(p)} = d_i \frac{\tau_{p}}{c} u_i^{(p)}
	\end{align*}
	from which the result unfolds. In the simulations on synthetic networks, the off-diagonal terms of the matrix $C$ are drawn from a uniform distribution $\mathcal{U}(c_{\rm out} - f,c_{\rm out} + f)$, the element $C_{11}$ is fixed to $c_{in}$ and all the other diagonal terms are determined to ensure $C\Pi\mathds{1}_k = c\mathds{1}_k$. The randomness will make the eigenvalues of $C\Pi$ non degenerate and there will not be a unique transition. The line $c_{\rm in} - c_{\rm out} = k\sqrt{c}$ indicates the approximated position of the transition.
	
\begin{figure}[h!]
		\centering
		\includegraphics[width = 0.6\columnwidth]{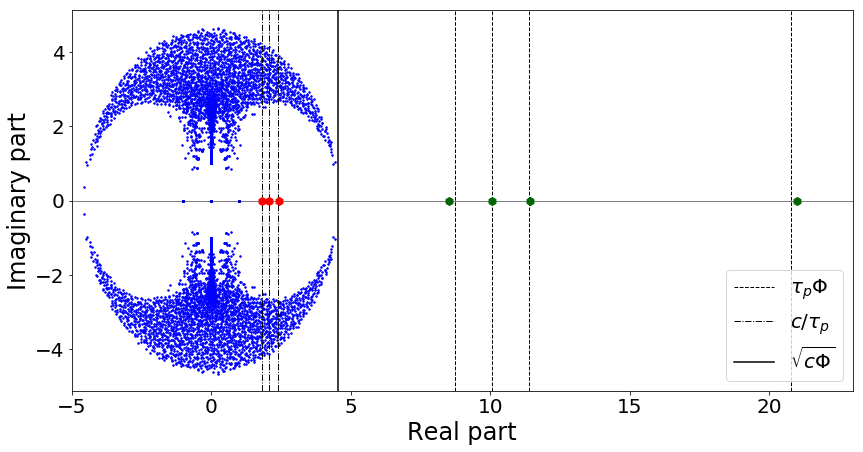}
		\caption{Spectrum of $B$. In green the isolated real eigenvalues outside the bulk corresponding to $\{\tau_p \Phi \}$, in red those inside the bulk, corresponding to $\{\zeta_p = c/\tau_p\}$; in blue all the others.
			We used $4$ clusters of equal size, $n = 5000$, $c_{\rm in} = 20$, $c_{\rm out} = 5$, $f = 1.5$ and $\theta_i \sim \theta = \mathcal{U}(3,13)^4$.}
		\label{fig:B_spec_more}
	\end{figure}

	In Figure~\ref{fig:B_spec_more} we report the spectrum of $B$ in the case of four classes, that shows that the largest isolated  real eigenvalues of the matrix $B$ are $\tau_p $ for $1 \leq p \leq  k$, followed by $c/\tau_p$ for $2 \leq p \leq k$. This result can be obtained analytically from the linearization of the belief propagation equations (see~\cite{krzakala2013spectral}).
	
\begin{figure}[h!]
		\centering
		\includegraphics[width=\columnwidth]{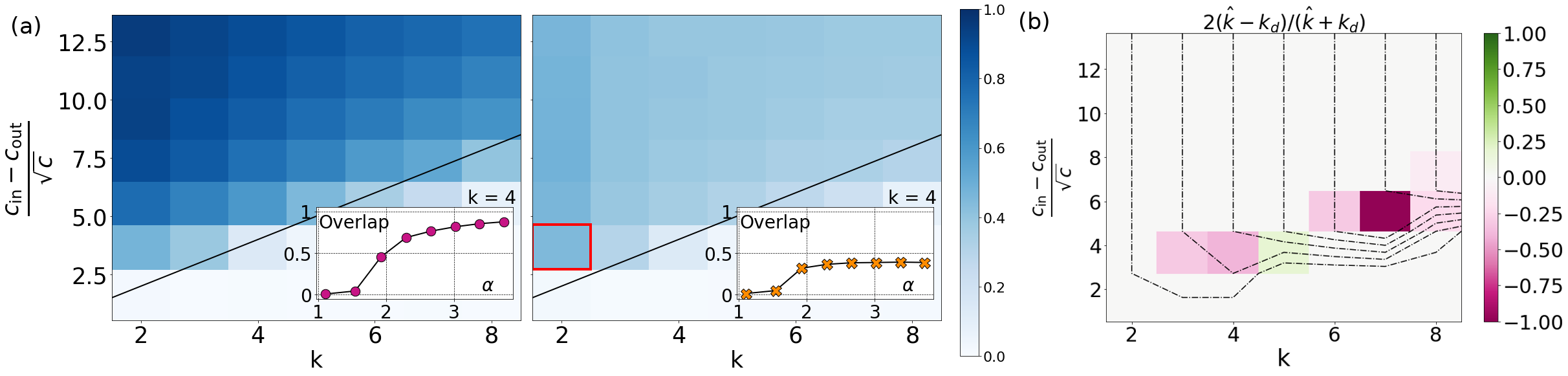}
		\caption{(a) Overlap (color scale) as a function of the number of classes ($k$) and hardness of the problem for the proposed algorithm (left) and $H_{\sqrt{c\Phi}}$ (right). Here, $n = 10$ $000$, $c_{\rm in} = 4 \to 40$, $c_{\rm out} = 3$, $f = 2/k$, $\theta_i \sim [\mathcal{U}(3,13)]^4$. Averaged over 10 samples.\\
			(b) Recovery ($2(\hat{k} - k_d)/(\hat{k} + k_d)$) as a function of $k$ and the hardness of the problem for the same parameters as (a).}
		\label{fig:more}
	\end{figure}

	Figure~\ref{fig:more}(a) displays the overlap as a function of the hardness of the problem and of the number of classes comparing our algorithm with \cite{saade2014spectral}, evidencing a strong advantage in terms of performance for our algorithm. The red square underlines the fact that the two methods coincide \emph{only} at the transition when $k = 2$ and the latter algorithm pays a lot in terms of performance for $k > 2$, even close to the transition. Figure~\ref{fig:more}(b) shows how $\hat{k} = |\{p,~v_p(\sqrt{c\Phi}) < 0\}|$ is a good estimator of the number of classes. With $k_d = |\{p,~\tau_p > \sqrt{c/\Phi}\}|$ we denote the number of theoretically detectable clusters and plot the quantity $2(\hat{k} - k_d)/(\hat{k} + k_d)$, showing small disagreement only close to the transition. The recovery being asymptotically exact, this can be interpreted as a finite size effect.

\end{document}